\documentclass[twocolumn]{aastex701}

\usepackage{graphicx,amsmath,amssymb}
\usepackage[export]{adjustbox}
\usepackage{booktabs}

\usepackage{siunitx}
\usepackage{makecell,caption}
\usepackage{listings}
\usepackage{xcolor}

\newcommand{\kms}{\ensuremath{\mathrm{km\,s}^{-1}}}

\usepackage{txfonts}

\usepackage{bm}
\usepackage[export]{adjustbox}
\usepackage{mathrsfs,amsmath,amssymb}

\def\mnras{Mon. Not. R. Astron. Soc. }

\def\apj{Astrophys. J.}
\def\apjl{Astrophys. J. Lett.}
\def\apjs{Astrophys. J., Suppl. Ser.}
\def\prd{Phys. Rev. D}

\def\aap{Astron. Astrophys.}

\begin{document}

\title{Taming the plunge: A circularization trap of supermassive black hole binaries}

\author[orcid=0000-0003-3993-3249,gname='Pau',sname='Amaro Seoane']{Pau Amaro Seoane}
\affiliation{Universitat Politècnica de València, Spain}
\affiliation{Max Planck Institute for Extraterrestrial Physics, Garching, Germany}
\email[show]{amaro@upv.es}

\author[orcid=0000-0002-2386-9142,gname='Alessandra',sname='Mastrobuono Battisti']{Alessandra Mastrobuono Battisti}
\affiliation{Dipartimento di Fisica e Astronomia ``Galileo Galilei'', Univ. di Padova, Padova, Italy}
\email[]{alessandra.mastrobuono@unipd.it}

\author[orcid=0000-0002-1672-894X,gname='Chingis',sname='Omarov']{Chingis Omarov}
\affiliation{Fesenkov Astrophysical Institute, Almaty 050020, Kazakhstan}
\email[]{chingis.omarov@fai.kz}

\author[orcid=0000-0002-5604-9757,gname='Denis',sname='Yurin']{Denis Yurin}
\affiliation{Fesenkov Astrophysical Institute, Almaty 050020, Kazakhstan}
\email[]{yurin@fai.kz}

\author[orcid='0000-0003-3643-9368',gname='Maxim',sname='Makukov']{Maxim Makukov}
\affiliation{Fesenkov Astrophysical Institute, Almaty 050020, Kazakhstan}
\email[]{makukov@fai.kz}

\author[orcid='0000-0002-5937-4985',gname='Dana',sname='Kuvatova']{Dana Kuvatova}
\affiliation{Fesenkov Astrophysical Institute, Almaty 050020, Kazakhstan}
\email[]{kuvatova@fai.kz}

\author[orcid='0000-0003-4782-8545',gname='Gulnara',sname='Omarova']{Gulnara Omarova}
\affiliation{Fesenkov Astrophysical Institute, Almaty 050020, Kazakhstan}
\email[]{gulnara.omarova@gmail.com}

\author[orcid='0000-0002-0738-7725',gname='Anton',sname='Gluchshenko']{Anton Gluchshenko}
\affiliation{Fesenkov Astrophysical Institute, Almaty 050020, Kazakhstan}
\email[]{gluchshenko@fai.kz}

\begin{abstract}
We investigate the orbital eccentricity evolution of supermassive black hole binaries (SMBHBs) ($10^4$--$10^7 M_{\odot}$), key sources for LISA, within realistic galactic environments. Merger timescales depend critically on eccentricity when gravitational wave (GW) emission dominates. We analyze the dynamics in triaxial merger remnants and subsequent interactions with geometrically thick nuclear discs (NSDs/CNDs). We confirm that gravitational torques in triaxial potentials efficiently extract angular momentum ($T_J \ll T_E$), resulting in binary formation with high initial eccentricities ($e > 0.95$). We then analyze the binary-disc interaction using a 3D analytical framework incorporating the Airy formalism and potential softening. We present a self-consistent derivation demonstrating that the 3D suppression of high-order torques (torque cutoff) leads to distinct scalings with disc thickness ($h$): migration rates $\tau_a^{-1} \propto h^{-3}$ and eccentricity damping rates $\tau_e^{-1} \propto h^{-5}$. This establishes a robust timescale hierarchy, $\tau_e/\tau_a \propto h^2$. For typical parameters ($h\approx 0.2$), eccentricity damping is significantly faster than orbital decay ($\tau_e \approx 0.04 \, \tau_a$). {We further develop a wavelet-based formalism to quantify the impact of disc inhomogeneities arising from accretion feedback and turbulence. We derive the stochastic torque variance in the wavelet domain and employ a Fokker-Planck analysis to determine the equilibrium eccentricity distribution. We demonstrate that while stochastic fluctuations counteract deterministic damping, the strong damping imposed by the thick disc geometry ensures the equilibrium eccentricity remains small unless the fluctuations are highly non-linear.} Consequently, while born highly eccentric, SMBHBs are rapidly circularized by the nuclear environment. This ``circularization trap'' forces binaries to approach the GW-dominated regime on nearly circular orbits, prolonging the total merger timescale. This introduces a substantial cosmological delay (0.3--2.5 Gyr) governed by stellar relaxation, which impacts LISA detection rates and the modeling of SMBH assembly in cosmological frameworks.
\end{abstract}


\section{Introduction}
\label{sec:introduction}

The hierarchical paradigm of galaxy formation posits that galaxies assemble through successive mergers. This process inevitably leads to the formation of supermassive black hole binaries (SMBHBs) at the centers of the resulting remnants \citep{BegelmanEtAl80, ColpiDotti2011}. The eventual coalescence of these binaries generates low-frequency gravitational waves (GWs). Binaries within the mass range $10^4-10^7\,M_{\odot}$ are anticipated to be primary sources for the Laser Interferometer Space Antenna (LISA) mission \citep{LISA2017}.

The dynamical evolution of an SMBHB spans an immense range of scales, commencing at kiloparsec separations following the galaxy merger and culminating at the milliparsec scale, where the emission of GWs becomes the dominant mechanism driving the binary towards coalescence. The efficiency of interactions between the binary and its environment---comprising both stars and gas---determines the overall timescale for the merger. Should these environmental mechanisms prove inefficient at extracting energy and angular momentum, the binary's evolution may stall, a scenario known as the final parsec problem \citep{MM03b}.

A parameter of paramount importance in this evolution is the orbital eccentricity, $e$. The timescale for coalescence driven by GW emission exhibits a steep dependence on eccentricity, as derived by \citet{Peters64}:

\begin{equation}
\tau_{\text{GW}} \propto a^4 (1-e^2)^{7/2},
\label{eq:tau_gw}
\end{equation}

\noindent
where $a$ is the semi-major axis. Consequently, binaries maintaining high eccentricity merge significantly faster than their circular counterparts at the same separation. Understanding the environmental processes that shape the eccentricity prior to the GW-dominated regime is therefore crucial for predicting LISA event rates.

The environment encountered by the SMBHs during this process is structurally complex. Galaxy mergers typically produce non-spherical (often triaxial) remnants and simultaneously drive substantial gas inflows toward the galactic center. This inflow facilitates the accumulation of mass in dense, rotationally supported structures known variously as Circumnuclear Discs (CNDs) or, when predominantly stellar, Nuclear Stellar Discs (NSDs). NSDs are massive, dense, flattened systems that frequently dominate the gravitational potential in the central few parsecs (see \citealt{SchultheisEtAl2025}). {These structures often coexist with Nuclear Star Clusters (NSCs), which are dense, spheroidal stellar systems \citep{NeumayerEtAl2020}. NSCs play a significant role in the dynamics; their high stellar density can accelerate the initial pairing of the SMBHs through enhanced dynamical friction. Furthermore, the dynamical interplay during the merger can alter the structure of the NSC, often leading to flattening and the acquisition of rotation \citep{HartmannEtAl2011,TsatsiEtAl2017}.} They are typically dynamically warm and geometrically thick, characterized by an aspect ratio $h=H/R \approx 0.1-0.3$. {Furthermore, accretion processes and the associated feedback mechanisms can drive turbulence and significant inhomogeneities within these structures.}

In this paper, we present a self-consistent analytical framework {\citep[building upon foundational work in gravitational dynamics and wave-disc interactions, e.g.,][]{GoldreichTremaine1980, Ward1986, TanakaEtAl2002}} to trace the evolution of eccentricity for SMBHBs relevant to the LISA band. We analyze the effects of the large-scale non-spherical potential during the initial sinking phase and provide a detailed analytical derivation of the subsequent interaction with the rotating, geometrically thick nuclear disc environment. We demonstrate that while binaries are likely born with high eccentricity, the fundamental physics of wave-disc interactions in thick media imposes rapid deterministic circularization. {We subsequently introduce a wavelet analysis to quantify how stochastic density fluctuations, potentially arising from accretion feedback, compete with this deterministic evolution.}

\section{Eccentricity at Binary Formation}
\label{sec:formation_eccentricity}

We analyze the dynamics of two SMBHs as they sink toward the center of a galaxy merger remnant, prior to forming a bound Keplerian binary. {The formalism employed here relies on established principles of galactic dynamics, specifically comparing the timescales for energy evolution (driven by dynamical friction) and angular momentum evolution (driven by gravitational torques). This approach is standard in the literature and characterizes the evolution based on the system's parameters, such as the masses, density profile, and degree of asymmetry of the remnant.} {This phase involves dynamics on scales ranging from kiloparsecs down to tens of parsecs within the larger merger remnant.} We adopt a reference equal-mass system with $M_1 = M_2 = 10^6 M_{\odot}$ (total mass $M_{\text{bin}} = 2 \times 10^6 M_{\odot}$), situated within a host environment characterized by a stellar velocity dispersion $\sigma \approx 100$ \kms.

\subsection{The Timescale of Gaseous Dynamical Friction}

We first assess the efficiency of Gaseous Dynamical Friction (GDF) during the approach phase. {Estimating this timescale is necessary to establish the hierarchy of dynamical processes. As shown below, friction timescales (both stellar and gaseous) are significantly longer than the timescale for angular momentum change via torques, which dictates the eccentricity evolution during this phase.} GDF arises from the gravitational wake induced by the moving SMBH in the surrounding gas. The timescale for orbital decay due to GDF is given by the linear theory analysis of \citet{Ostriker1999}:
\begin{equation}
\tau_{\text{GDF}} \approx \frac{V_c^3}{4\pi G^2 M_{\bullet} \rho_{\text{gas}} C},
\label{eq:tau_gdf}
\end{equation}
where $V_c$ is the circular velocity {(used here as a proxy for the characteristic relative velocity between the SMBH and the background medium, which is appropriate for the sinking phase where orbits are often eccentric or misaligned)}, $M_{\bullet}$ the SMBH mass, $\rho_{\text{gas}}$ the ambient gas density, and $C$ is a dimensionless factor dependent on the Mach number of the perturber.

Adopting representative parameters ($V_c = 150$ \kms, $\rho_{\text{gas}} = 10 M_{\odot}/\text{pc}^3$, $C = 3$), we find $\tau_{\text{GDF}} \approx 473$ Myr. This timescale is substantially longer than the local dynamical time (approximately 6.5 Myr at a radius of 1 kpc). {While this process contributes to the overall orbital decay, the crucial metric for determining the orbital evolution is the comparison with other dynamical mechanisms. As demonstrated below, gravitational torques operate on a much shorter timescale. Therefore, GDF (and similarly, stellar dynamical friction) is subdominant in shaping the orbit during the large-scale sinking phase for these masses.}

\subsection{Dominance of Gravitational Torques in Non-Spherical Potentials}
\label{sec:gravitational_torques}

The dominant factor shaping the orbits of the sinking SMBHs is the non-spherical morphology of the background gravitational potential, which arises naturally from the violent relaxation processes during the merger. We characterize the degree of asymmetry by a dimensionless parameter $\epsilon$. {This parameter quantifies the deviation from sphericity, encompassing triaxiality or strong non-axisymmetric features (e.g., bars). These features are crucial because they break the conservation of angular momentum components (such as $L_z$ in axisymmetric systems) and thereby allow significant changes in orbital shape.}

To understand the orbital evolution, it is crucial to compare the timescale for angular momentum change ($T_J$) with the timescale for energy change ($T_E$). {We assume a static or slowly rotating non-spherical potential; in a rapidly rotating or strongly time-dependent potential, energy is not conserved even without friction, complicating the analysis, though the dominance of torques often persists.}

In this context, $T_E$ is governed by dynamical friction (DF). {DF represents the change in orbital energy due to the cumulative effect of weak gravitational encounters. The instantaneous rate of energy change is $\dot{E} = \vec{F}_{\text{DF}} \cdot \vec{v}$. It is noted that the effect of DF in coherently rotating backgrounds is complex; friction can increase energy at apocenters, potentially leading to suppressed net energy loss or quasi-conserved energy during circularization \citep[e.g.][]{BonettiEtAl2020}.}

{However, to establish the timescale hierarchy, we require a characteristic timescale $T_E$ representing the magnitude of the energy exchange. The standard dynamical friction timescale, $\tau_{DF}$, is derived from the Fokker-Planck formalism or related approaches and quantifies the magnitude of the drag force $||\vec{F}_{\text{DF}}||$. Conventionally, $\tau_{DF}$ is defined as the timescale for the velocity to change by order itself, i.e., $\tau_{DF} \approx M_{\bullet} V_c / ||\vec{F}_{\text{DF}}||$. We define the energy timescale based on the characteristic magnitude of the energy change rate:}

\begin{equation}
T_E \equiv \frac{|E|}{\langle |\dot{E}| \rangle} \approx \frac{M_{\bullet} V_c^2}{2 \langle \|\vec{F}_{\mathrm{DF}}\| V_c \rangle}.
\label{eq:T_E_definition}
\end{equation}

\noindent {It follows dimensionally that $T_E \approx \tau_{DF}$. This identification is mathematically justified because $\tau_{DF}$ quantifies the fundamental rate of the diffusive interaction. Furthermore, in a pressure-supported triaxial remnant where velocity dispersion dominates rotation, the standard formulation (Eq.~\ref{eq:T_E_analytical}) is robust. Even if rotation effects were significant enough to suppress energy loss ($T_E > \tau_{DF}$), using $\tau_{DF}$ provides a conservative lower bound for $T_E$, which would only strengthen the argument that $T_J \ll T_E$.} Conversely, $T_J$ represents the timescale over which coherent gravitational torques, arising from the large-scale asymmetries ($\epsilon$), modify the orbital angular momentum.

The shape of the orbit (its eccentricity) is determined by the relative efficiency of these two processes. If $T_E \ll T_J$, energy changes while angular momentum is approximately conserved (in a near-spherical system), typically leading to circularization. If $T_J \ll T_E$, angular momentum is extracted faster than energy is dissipated by friction, driving the orbit towards a radial trajectory (high eccentricity).

We can estimate $T_J$ as follows. The radial gravitational force is approximately $F_{\text{rad}} \approx M_{\bullet} V_c^2/R$. The tangential force arising from the non-spherical perturbation is $F_{\text{tan}} \sim \epsilon F_{\text{rad}}$. The resulting torque is $\tau \sim R F_{\text{tan}} \sim \epsilon M_{\bullet} V_c^2$. {We define $L_{max} \approx M_{\bullet} R V_c$ as the characteristic maximum angular momentum at radius R (corresponding to a circular orbit) to set the scale for $T_J$.} The timescale for angular momentum change is:
\begin{equation}
T_J \sim \frac{L_{max}}{\tau} \sim \frac{M_{\bullet} R V_c}{\epsilon M_{\bullet} V_c^2} = \frac{R}{\epsilon V_c} = \frac{T_{\text{dyn}}}{\epsilon}.
\label{eq:T_J_analytical}
\end{equation}
\noindent {This derivation follows standard timescale arguments used in galactic dynamics for estimating the efficiency of gravitational torques \citep[see, e.g.,][]{BinneyTremaine08}.} This reveals that the timescale for angular momentum evolution via torques is comparable to the dynamical time, modulated only by the strength of the asymmetry. {For realistic asymmetries ($\epsilon \approx 0.1-0.3$), $T_J$ is only a few times $T_{\text{dyn}}$.}

The energy change timescale is $T_E = \tau_{\text{DF}}$. Assuming an isothermal sphere density profile, $\rho(R) = V_c^2 / (4\pi G R^2)$, and utilizing the enclosed mass $M_{\text{enc}}(R) = V_c^2 R / G$, we can express $T_E$ as: {The assumption of an isothermal sphere ($\rho \propto r^{-2}$) is a standard approximation for the overall mass distribution in pressure-supported systems like elliptical galaxies and merger remnants. While galaxy luminosity profiles often deviate from this (e.g., steepening towards the center), the isothermal model provides a robust estimate for the dynamical friction timescale across the scales (kpc to 100 pc) relevant for this sinking phase, as the timescale primarily depends on the local density and velocity dispersion.}
\begin{align}
T_E &\approx \frac{V_c^3}{4\pi G^2 M_{\bullet} \rho C} = \frac{V_c R^2}{G M_\bullet C} \nonumber \\
&= T_{\text{dyn}} \frac{M_{\text{enc}}}{M_\bullet C}.
\label{eq:T_E_analytical}
\end{align}
\noindent {Here, $C$ is the dimensionless drag coefficient (e.g., the Coulomb logarithm for stellar DF or the Mach factor for GDF), which is typically of order unity or slightly larger.}

\noindent The ratio of these timescales is crucial:
\begin{equation}
\frac{T_J}{T_E} \approx \frac{M_\bullet C}{\epsilon M_{\text{enc}}}.
\label{eq:T_J_T_E_ratio_analytical}
\end{equation}

\noindent At a representative large radius $R=1$ kpc, $M_{\text{enc}} \approx 5.2 \times 10^9 M_{\odot}$. {This value represents the total dynamical mass enclosed within 1 kpc, derived from the assumed isothermal profile. This scale typically encompasses the bulge and lies well outside the NSD. The value strongly depends on the galaxy type; for comparison, the Milky Way's dynamical mass within 1 kpc is estimated to be around $1-2 \times 10^{10} M_{\odot}$ \citep{PortailEtAl2017}. Our adopted value is representative of an intermediate-mass merger remnant.} Using our reference parameters ($M_{\bullet} = 10^6 M_{\odot}$, $C=5$, $\epsilon=0.2$, {representing a typical triaxiality in merger remnants}), we find $T_J/T_E \approx 4.8\times 10^{-3}$. {This hierarchy persists as the SMBHs sink deeper, provided the environment remains non-spherical. For instance, at $R=100$ pc, which is a typical scale for the transition to the nuclear region, $M_{\text{enc}} \approx 5.2 \times 10^8 M_{\odot}$ (assuming the isothermal profile continues), yielding $T_J/T_E \approx 0.048$. This still ensures efficient angular momentum loss.}

The condition $T_J \ll T_E$ is robustly satisfied because the enclosed galactic mass remains substantially larger than the SMBH mass across the scales relevant to the sinking phase. This signifies a fundamental aspect of the dynamics: energy change via dynamical friction is a slow, diffusive process dependent on the mass ratio $M_\bullet/M_{\text{enc}}$. In contrast, gravitational torques provide a rapid, coherent mechanism for angular momentum transport dependent only on the geometry of the potential ($\epsilon$). Because the angular momentum is extracted much faster than the orbital energy is dissipated by friction, the SMBHs rapidly lose their tangential velocity. This forces them onto nearly radial trajectories during the sinking phase.

\subsection{Initial Eccentricity Estimation}

{As the SMBHs sink from larger scales (kpc to hundreds of parsecs) towards the center, they eventually form a bound binary.} The binary typically forms near the hardening radius $a_h = G M_{\text{bin}}/(4\sigma^2)$. {Concurrently, gas inflows driven by the merger also experience gravitational torques, losing angular momentum and migrating toward the center.}

The radius of influence of the binary is $R_{\text{inf}} = G M_{\text{bin}}/\sigma^2 \approx 0.86$ pc for our reference parameters. {$R_{\text{inf}}$ defines the characteristic scale where the binary's gravity begins to dominate the background potential.}

{The gas will settle into a rotationally supported structure at the circularization radius, $R_{\text{circ}}$, determined by its residual specific angular momentum. The physical reason why $R_{\text{circ}}$ is typically located near or within $R_{\text{inf}}$, rather than at much larger radii, is the efficient angular momentum transport occurring during the inflow. At large radii, the strong triaxiality of the merger remnant exerts significant gravitational torques on the gas (similar to the process affecting the SMBHs), driving the inflow. This process continues until the gas approaches $R_{\text{inf}}$, where the potential geometry transitions. Near the center, the potential becomes dominated by the central binary and the stellar cusp, often becoming more axisymmetric or spherical than the large-scale remnant. This change in symmetry significantly reduces the efficiency of the coherent gravitational torques, allowing the gas to settle into equilibrium at a radius determined by its remaining angular momentum.}

We parameterize this scale relative to $R_{\text{inf}}$. {We define $\beta$ as the fraction of the maximum (circular) specific angular momentum retained by the inflowing material at $R_{\text{inf}}$.} Assuming 50\% angular momentum retention ($\beta=0.5$) for the inflowing material relative to the maximum (circular) angular momentum at $R_{\text{inf}}$, the characteristic circularization radius where the disc forms is $R_{\text{circ}} = \beta^2 R_{\text{inf}} \approx 0.22$ pc. We adopt $a \approx 0.22$ pc as the initial separation of the newly formed binary, embedded within this nascent disc.

The eccentricity is defined by $e = \sqrt{1 - j^2}$, where $j=L/L_{\text{circ}}$ is the circularity. Due to the efficient angular momentum extraction described above, the SMBHs approach the center on highly radial paths. The residual tangential velocity acquired over the final dynamical time due to the asymmetric forces can be estimated as $V_t \sim F_{\text{tan}} T_{\text{dyn}}/M_\bullet \sim (\epsilon V_c^2/R) (R/V_c) \sim \epsilon V_c$. Consequently, the circularity just prior to binding is $j \sim V_t/V_c \sim \epsilon$.

Assuming a typical triaxiality in the merger remnant $\epsilon \approx 0.1-0.3$, the resulting eccentricity is $e \approx \sqrt{1-\epsilon^2}$, ranging from $0.995$ to $0.954$. This analysis confirms that the efficient angular momentum extraction during the large-scale sinking phase, facilitated by the underlying dynamics of {non-axisymmetric (triaxial)} systems which robustly support centrophilic orbits \citep{PoonMerritt01,PoonMerritt2004,VasilievMerritt2013}, results in binaries forming with characteristically high initial eccentricity, $e_{\text{initial}} > 0.95$.

\section{The Circularization Trap in Nuclear Discs}
\label{sec:circularization_trap}

The newly formed, highly eccentric binary is embedded within the dense, rotating environment of the NSD or CND. These structures possess significant internal energy (manifested as pressure in gaseous discs or velocity dispersion in stellar discs), resulting in a geometrically thick configuration. We adopt a reference aspect ratio $h = H/R \approx 0.2$.

{The analysis presented in this section relies on the framework of Type-I migration, which assumes that the binary does not open a deep gap in the disc. This is appropriate if the disc is sufficiently viscous or thick. If a gap is opened (Type-II migration), the dynamics change significantly; specifically, interactions at the gap edge and the suppression of corotation torques can potentially drive eccentricity growth rather than damping. We proceed under the assumption that the conditions in the dense nuclear environment (e.g., turbulence or high velocity dispersion) prevent the formation of a deep gap.}

{We analyze the evolution by considering two components: the deterministic evolution driven by the mean disc structure, and the stochastic evolution driven by inhomogeneities.}

\subsection{Deterministic Evolution: 3D Wave Dynamics}
\label{sec:airy_formalism}

The interaction between the SMBHB and the mean structure of the nuclear disk (NSD/CND) is mediated by the excitation of density waves at Lindblad resonances. These resonances occur at locations where the Doppler-shifted frequency of the binary's gravitational potential matches the natural epicyclic frequency of the disk material. While the WKB approximation is often sufficient for analyzing infinitesimally thin, cold disks, NSDs/CNDs are geometrically thick ($h \approx 0.2$) and possess significant internal energy. This internal energy fundamentally alters the wave dynamics and invalidates the standard WKB approach near resonances.

We employ a treatment combining the Airy formalism \citep{Ward1986} with three-dimensional (3D) corrections \citep{TanakaEtAl2002}. This framework is necessary because it captures the essential physics of wave launching in a pressurized, finite-thickness medium where the assumptions underlying WKB break down.

The failure of WKB stems from neglecting pressure gradients. As shown below in the dispersion relation (Eq. \ref{eq:dispersion_relation}), the radial wavenumber $k_r$ approaches zero near the resonance. This location acts as a classical turning point in the wave equation. Without pressure, the predicted wavelength would become arbitrarily small, leading to a singularity in the torque calculation. Physically, pressure counteracts gravitational compression, smoothing the disk's response.

\subsubsection{Derivation and Interpretation of the Forced Wave Equation}

We analyze the response of the disc to a component of the forcing potential $\Phi_m(R) e^{i(m\theta - \omega t)}$, where $\omega = m\Omega_p$ is the pattern speed (forcing frequency). {Here, $\Omega_p$ is the pattern speed of the specific potential component, which for a binary is related to its orbital frequency.} {The external potential $\Phi_m$ represents the $m$-th azimuthal harmonic of the binary's gravitational field.} The dynamics of the disk perturbation are derived from the linearized fluid equations. These represent the conservation of momentum and mass for small deviations from the background equilibrium flow. Let $u'$ and $v'$ be the radial and azimuthal velocity perturbations, respectively. The equations of motion are:

\begin{align}
i(\omega - m\Omega) u' - 2\Omega v' &= -\frac{\partial \Psi'}{\partial R} - \frac{\partial \Phi_m}{\partial R}, \label{eq:EOM_radial} \\
i(\omega - m\Omega) v' + \frac{\kappa^2}{2\Omega} u' &= -\frac{im}{R} \Psi' - \frac{im}{R} \Phi_m, \label{eq:EOM_azimuthal}
\end{align}
where $\Psi'$ is the perturbed enthalpy (related to pressure; we assume an isothermal equation of state with sound speed $c_s$), $\Omega(R)$ is the disc angular velocity, and $\kappa(R)$ is the epicyclic frequency. The velocity perturbations ($u', v'$) are driven by gradients in both the perturbed enthalpy ($\Psi'$) and the external binary potential ($\Phi_m$).

Following the procedure outlined by \citet{Ward1986}, based on the foundational theory of \citet{GoldreichTremaine1980}, we combine these equations to eliminate the velocity perturbations, arriving at a single second-order differential equation for the enthalpy perturbation:

\begin{equation}
\frac{d^2\Psi'}{dR^2} + k_r^2(R) \Psi' = S(R).
\label{eq:forced_wave_equation}
\end{equation}

\noindent This is a forced wave equation, mathematically analogous to a forced harmonic oscillator. It provides a clear physical interpretation of the interaction.

The left-hand side represents the homogeneous wave equation. It describes the natural, free propagation of density waves within the disk. The term $k_r^2(R)$ is the radial wavenumber squared, defined by the dispersion relation:
\begin{equation}
k_r^2(R) c_s^2 = m^2(\Omega(R)-\Omega_p)^2 - \kappa^2(R) \equiv D(R).
\label{eq:dispersion_relation}
\end{equation}
\noindent This relation encapsulates the intrinsic restoring forces of the disk medium: pressure gradients (via $c_s^2$), Coriolis forces, and differential rotation (via $\Omega$ and $\kappa$).

The right-hand side, $S(R)$, is the forcing term. It represents the external driver---the gravitational influence of the orbiting binary. $S(R)$ is explicitly defined by the structure and gradients of the binary's potential component $\Phi_m$. It takes a form involving derivatives of $\Phi_m$ weighted by the disk properties:

\begin{equation}
S(R) \approx \frac{1}{c_s^2} \left[ D(R) \Phi_m - c_s^2 \frac{d^2\Phi_m}{dR^2} + \mathcal{O}\left(\frac{d\Phi_m}{dR}\right) \right].
\end{equation}

\noindent Physically, the term "forced" signifies that the binary's gravitational potential is actively driving the disk material away from equilibrium. $S(R)$ acts as the source term that excites the waves, continuously inputting energy and angular momentum into the disk by driving its natural modes of oscillation. The forced wave equation describes how the amplitude of the resulting wave ($\Psi'$) is determined by the balance between the disk's intrinsic propagation properties ($k_r^2(R)$) and the strength of the external gravitational forcing ($S(R)$).

\subsubsection{The Airy Formalism: Regularization at Resonance}

The Lindblad resonance $R_L$ occurs where $D(R_L)=0$. Near the resonance, $k_r^2 \to 0$, corresponding to the turning point where the WKB approximation fails. We must solve Eq.~(\ref{eq:forced_wave_equation}) directly in this vicinity.

The Airy formalism is the mathematical technique required to solve the wave equation near this turning point. We introduce a local coordinate $x = R - R_L$ and Taylor expand the discriminant $D(x) \approx x D'$. Assuming the forcing term is slowly varying ($S(R) \approx S(R_L)$), Eq.~(\ref{eq:forced_wave_equation}) locally becomes:
\begin{equation}
\frac{d^2\Psi'}{dx^2} + \frac{D'}{c_s^2} x \Psi' = S(R_L).
\label{eq:wave_equation_linearized}
\end{equation}

\noindent By incorporating the pressure term $c_s^2$, the formalism transforms the governing differential equation locally into the inhomogeneous Airy equation. This transformation naturally defines a characteristic pressure length scale, $\Delta_p$:
\begin{equation}
\Delta_p^3 = \frac{c_s^2}{|D'|}.
\label{eq:Delta_p_definition}
\end{equation}
\noindent This scale defines the radial extent over which the wave behavior is regularized by pressure gradients. By introducing a scaled coordinate $\xi \propto x/\Delta_p$, Eq.~(\ref{eq:wave_equation_linearized}) transforms into the standard Airy form:

\begin{equation}
\frac{d^2 \Psi'}{d\xi^2} - \xi \Psi' = S_{\text{scaled}}.
\label{eq:Airy_equation}
\end{equation}

\noindent The solutions, expressed as Airy functions, physically describe the transition of the density perturbation from an oscillatory wave in the propagation zone ($k_r^2>0$) to an exponentially decaying (evanescent) response across the resonance. This formalism ensures that the density perturbations remain finite, resolving the WKB singularity \citep{GoldreichTremaine1980}.

\subsubsection{3D Corrections and the Torque Cutoff}
\label{sec:torque_cutoff}

Furthermore, the finite thickness $H$ of the disc necessitates 3D corrections. The gravitational potential of the binary, which orbits in the midplane, must be averaged over the vertical extent of the disk to determine the effective force experienced by the bulk of the disc material.

We analyze the 3D interaction, as formalized by \citet{TanakaEtAl2002}. Assuming a vertically isothermal disc, the density profile is $\rho(R,z) = \rho_0(R) \exp(-z^2/2H^2)$. The effective potential component driving the wave, $\Phi_m^{\text{eff}}$, is obtained by a convolution of the 3D potential with the vertical density structure:
\begin{equation}
\Phi_m^{\text{eff}}(R) = \frac{1}{\sqrt{2\pi}H} \int_{-\infty}^{\infty} \Phi_m^{3D}(R,z) \exp\left(-\frac{z^2}{2H^2}\right) dz.
\label{eq:3d_potential_averaging}
\end{equation}

\noindent This integration reveals a significant weakening of the effective potential compared to the midplane potential when the characteristic radial wavelength of the forcing potential, $\lambda_F \sim R/m$, becomes comparable to or smaller than the disc scale height $H$. This phenomenon is termed potential softening. The physical interpretation is straightforward: when the forcing varies rapidly in the radial direction (high $m$), the vertical averaging process smooths out these variations, reducing the net force driving the wave.

\citet{TanakaEtAl2002} quantified this suppression by deriving the softening factor $f_s = \Phi_m^{\text{eff}}/\Phi_m^{\text{midplane}}$. An analytical approximation is given by:
\begin{equation}
f_s(mh) \approx \left[ 1 + \left(\frac{0.876}{0.4}\right)^2 (mh)^2 \right]^{-3/4}.
\label{eq:Tanaka_softening}
\end{equation}
\noindent This demonstrates a strong suppression of torques when the dimensionless parameter $mh \gtrsim 1$. This leads to a ``torque cutoff'' at high azimuthal numbers. The cutoff occurs approximately when $\lambda_F \approx H$, corresponding to:
\begin{equation}
m_{\text{cut}} \approx \frac{R}{H} = \frac{1}{h}.
\label{eq:m_cut}
\end{equation}
\noindent For a typical thickness $h=0.2$, $m_{\text{cut}} \approx 5$. This suppression is essential for calculating the correct torques in a realistic thick disk environment, and, as shown below, it disproportionately affects the evolution of eccentricity, which relies heavily on these high-$m$ modes.

\subsection{Deterministic Timescale Hierarchy}
\label{sec:analytical_derivation}

We now present an analytical derivation of the scaling of the migration timescale ($\tau_a$) and the eccentricity damping timescale ($\tau_e$) with the disc thickness $h$, explicitly incorporating the 3D effects discussed above. This derivation illuminates the physical origin of the timescale hierarchy in the laminar disc approximation.

\subsubsection{Modeling 3D Suppression}

We incorporate the torque suppression by introducing a factor $S(m, h)$ to the idealized two-dimensional (2D) WKB torque $\Gamma_m^{2D}$:
\begin{equation}
\Gamma_m^{3D} = \Gamma_m^{2D} S(m, h).
\end{equation}
\noindent The suppression factor $S(m,h)$ contains the combined effects of 3D potential softening and the modification of wave propagation in the vertical direction. The detailed 3D analysis \citep{TanakaEtAl2002} yields a suppression factor that depends primarily on the dimensionless quantity $x = mh$. A representative functional form capturing the behavior derived from their analysis is:
\begin{equation}
S(x) \approx \left(1 + \left(\frac{x}{x_c}\right)^2\right)^{-3/2},
\label{eq:suppression_factor_3D}
\end{equation}
where $x_c$ is a constant of order unity. Crucially, for $x \gg 1$ (i.e., $m \gg m_{\text{cut}}$), the suppression factor scales as $S(x) \propto x^{-3}$. This steep dependence reflects the efficiency with which the vertical structure smooths out the high-frequency components of the gravitational potential.

\subsubsection{Scaling of the Migration Timescale ($\tau_a$)}

The migration rate (the {orbit-averaged} change in semi-major axis) is proportional to the total torque exerted on the binary. We approximate the summation over azimuthal modes $m$ by an integral:
\begin{equation}
\tau_a^{-1} \propto \int dm \, \Gamma_m^{2D} S(mh).
\label{eq:tau_a_integral}
\end{equation}

\noindent Migration is dominated by low-$m$ modes, corresponding to large-scale spiral waves. Assuming $\Gamma_m^{2D}$ is a slowly varying function of $m$ for these dominant modes, we perform a change of variables to $x=mh$:
\begin{equation}
\tau_a^{-1} \propto \int \frac{dx}{h} \, \Gamma^{2D} S(x) = \frac{1}{h} \left(\Gamma^{2D} \int S(x) dx\right).
\label{eq:tau_a_scaled}
\end{equation}

\noindent The mathematical justification for treating $\Gamma_m^{2D}$ as a slowly varying function of $m$ for the dominant modes relies on the interplay between the spectral properties of the gravitational potential and the sharp cutoff imposed by the 3D suppression factor. The torque $\Gamma_m^{2D}$ is proportional to the square of the amplitude of the forcing potential component, $|\Phi_m|^2$. For low azimuthal numbers $m$, the associated radial wavelength, $\lambda_F \sim R/m$, is large. In this long-wavelength regime, the Fourier components of the potential (related to Laplace coefficients) exhibit a smooth and relatively weak dependence on $m$, resulting in a characteristically flat torque spectrum $\Gamma_m^{2D}(m)$ at low $m$. Crucially, the integral for the total migration rate is sharply constrained by the suppression factor $S(mh)$, which decays rapidly (as $(mh)^{-3}$) for $mh \gtrsim 1$. This acts as a low-pass filter, effectively truncating the integral at $m \approx 1/h$. Since the dominant contribution to the integral is restricted to this low-$m$ regime where the variation of $\Gamma_m^{2D}$ is negligible compared to the rapid variation of $S(mh)$, we can approximate $\Gamma_m^{2D}$ as constant. This allows it to be factored out following the change of variables to $x=mh$, correctly isolating the dependence of the total torque on the disc thickness $h$.

The integral $\int S(x) dx$ converges rapidly to a constant value due to the $x^{-3}$ scaling of $S(x)$ at large $x$. Therefore, the 3D effects introduce an overall suppression factor of $h^{-1}$ relative to the 2D migration rate. The standard 2D Type I migration rate is known to scale as $\tau_a^{-1}(2D) \propto h^{-2}$. Substituting this yields the 3D scaling:
\begin{equation}
\tau_a^{-1}(3D) \propto h^{-2} \, h^{-1} = h^{-3}.
\label{eq:tau_a_scaling}
\end{equation}

\noindent {It is important to note that high eccentricity generally reduces the efficiency of migration compared to the circular case, as the interaction time at resonances is reduced and the relative velocity is higher. Therefore, the derived $\tau_a$ should be considered a lower limit for the migration timescale when the binary is still eccentric. A longer $\tau_a$ due to eccentricity would further reinforce the timescale hierarchy derived below.}

\subsubsection{Scaling of the Eccentricity Damping Timescale ($\tau_e$)}

The eccentricity damping rate depends on a weighted sum of the torques. Unlike migration, the evolution of eccentricity is significantly more sensitive to high-$m$ modes. This sensitivity arises because an eccentric orbit involves epicyclic motion, which couples strongly to higher harmonics of the potential that resonate further out or closer in. The contribution of these modes to eccentricity evolution is weighted more heavily than their contribution to migration. We model this dependence by assuming a power-law weighting, $W_m^e \propto m^2$, consistent with analyses of eccentric Lindblad resonances:
\begin{equation}
\tau_e^{-1} \propto \int dm \, m^2 \Gamma_m^{2D} S(mh).
\label{eq:tau_e_integral}
\end{equation}

\noindent {This derivation relies on the contribution of Lindblad resonances. Eccentricity evolution is also driven by corotation resonances, which typically contribute to damping. If the binary opens a gap, the material at corotation is depleted. This can drastically alter the eccentricity evolution, potentially leading to eccentricity growth rather than damping if the Lindblad torques are insufficient to overcome the effects of the depleted corotation region. Our analysis assumes the Type I regime where a significant gap is not present.}

Again, we change variables to $x=mh$:
\begin{equation}
\tau_e^{-1} \propto \int \frac{dx}{h} \left(\frac{x}{h}\right)^2 \Gamma^{2D} S(x) = \frac{1}{h^3} \left(\Gamma^{2D} \int x^2 S(x) dx\right).
\label{eq:tau_e_scaled}
\end{equation}
\noindent The integral $\int x^2 S(x) dx$ also converges to a constant. Consequently, the 3D suppression factor for eccentricity damping is $h^{-3}$. This suppression is much stronger than the $h^{-1}$ factor found for migration, because the $m^2$ weighting emphasizes the high-$m$ modes which are severely affected by the torque cutoff.

Assuming the underlying 2D eccentricity damping rate scales similarly to the migration rate ($\tau_e^{-1}(2D) \propto h^{-2}$), we obtain the 3D scaling:
\begin{equation}
\tau_e^{-1}(3D) \propto h^{-2} \cdot h^{-3} = h^{-5}.
\label{eq:tau_e_scaling}
\end{equation}

\subsubsection{The Timescale Hierarchy}

By comparing the analytically derived scalings (Eq.~(\ref{eq:tau_a_scaling}) and Eq.~(\ref{eq:tau_e_scaling})), we establish the fundamental timescale hierarchy:
\begin{equation}
\frac{\tau_e}{\tau_a} = \frac{\tau_a^{-1}}{\tau_e^{-1}} \propto \frac{h^{-3}}{h^{-5}} = h^2.
\end{equation}
\noindent This leads to the relation:
\begin{equation}
\frac{\tau_e}{\tau_a} \approx P h^2,
\label{eq:tau_ratio_Airy}
\end{equation}
\noindent where $P$ is a dimensionless factor of order unity.

The physical interpretation of this $h^2$ scaling is critical. It demonstrates that the efficiency of eccentricity damping relative to migration decreases quadratically as the disc thickness increases. However, provided $h \ll 1$, the condition $\tau_e \ll \tau_a$ remains satisfied. The derivation confirms that rapid circularization is an inevitable consequence of the interaction in the Type I regime, driven by the differential suppression of high-$m$ (eccentricity-driving) modes compared to low-$m$ (migration-driving) modes in a thick disc.

{It is instructive to contrast these analytical findings with results from 3D numerical simulations, some of which report different scaling relations or slower circularization rates. The apparent discrepancies primarily arise from differences in the physical regimes being studied rather than fundamental disagreements about the underlying wave mechanics.}

{The analytical framework presented here is derived within the linear Type-I regime, applicable to thick discs ($h \approx 0.2$) where pressure support prevents gap opening. The derived scalings ($\tau_a^{-1} \propto h^{-3}$, $\tau_e^{-1} \propto h^{-5}$) are robust mathematical consequences of the 3D geometry and the associated torque cutoff (Eq.~\ref{eq:suppression_factor_3D}).}

{Conversely, numerical simulations often investigate regimes where non-linear effects or gap formation (Type-II migration) become significant. This occurs particularly in simulations with thinner discs (lower $h$) or higher binary mass ratios ($q$), which facilitate gap opening. The transition to the Type-II regime fundamentally alters the dynamics. Gap formation depletes the material at corotation resonances, suppressing the damping effect of these torques. Simultaneously, interactions at the gap edges can become dominant, often leading to eccentricity excitation rather than damping.}

{Furthermore, the behavior of corotation torques is complex even without a full gap. Linear theory assumes these torques operate fully. However, they can saturate (weaken over time) if the gradients of vortensity and entropy are erased faster than they can be replenished by diffusion. This saturation depends sensitively on the disc's thermodynamic properties and viscosity prescription, which vary significantly between simulations.}

{Therefore, slower circularization observed in some simulations typically reflects the onset of Type-II dynamics, saturation effects, or strong non-linear dissipation, rather than contradicting the linear 3D theory. The analytical model accurately describes the evolution in the specific context of thick nuclear discs considered here, where the Type-I approximation is justified. The value of this analytical approach lies in isolating the baseline behavior and the fundamental scalings imposed by the disc geometry in this specific regime.}

\subsection{{Stochastic Evolution: Wavelet Analysis of Feedback-Induced Inhomogeneities}}
\label{sec:wavelet_analysis}

{The preceding analysis addresses the deterministic evolution driven by the mean disc structure. However, accretion feedback and turbulence introduce stochastic density fluctuations, which exert fluctuating torques on the binary. These fluctuations can counteract the deterministic damping. To quantify this stochastic excitation across the relevant spatial scales, we employ a formal wavelet analysis.}

\subsubsection{{Wavelet Decomposition of the Stochastic Density Field}}

{We model the surface density as $\Sigma(\mathbf{x}, t) = \Sigma_0(\mathbf{x}) + \delta\Sigma(\mathbf{x}, t)$, where $\delta\Sigma(\mathbf{x}, t)$ is the fluctuating component with $\langle \delta\Sigma \rangle = 0$. We decompose $\delta\Sigma$ onto an orthonormal wavelet basis $\{\psi_{\lambda}(\mathbf{x})\}$, which spans the Hilbert space $L^2(\mathcal{D})$ over the disc domain $\mathcal{D}$. The multi-index $\lambda = (j, \mathbf{k})$ denotes the scale ($s_j = 2^{-j}$) and location ($\mathbf{x}_{\mathbf{k}} = 2^{-j}\mathbf{k}$). The decomposition is formally given by:}
\begin{equation}
\delta\Sigma(\mathbf{x}, t) = \sum_{\lambda \in \Lambda} c_{\lambda}(t) \psi_{\lambda}(\mathbf{x}),
\label{eq:wavelet_decomposition_density}
\end{equation}
\noindent {where the wavelet coefficients $c_{\lambda}(t) = \langle \delta\Sigma(t), \psi_{\lambda} \rangle = \int_{\mathcal{D}} \delta\Sigma(\mathbf{x}, t) \psi_{\lambda}^*(\mathbf{x}) d^2\mathbf{x}$ quantify the localized amplitude of the fluctuation at scale and position $\lambda$.}

\subsubsection{{Stochastic Torque Formalism}}

{The stochastic component of the torque, $\Gamma_{\text{stoch}}(t)$, arises from the interaction of the binary potential $\Phi_{\text{bin}}$ with the density fluctuations:}
\begin{equation}
\Gamma_{\text{stoch}}(t) = -\int_{\mathcal{D}} \delta\Sigma(\mathbf{x}, t) \nabla_{\theta} \Phi_{\text{bin}}(\mathbf{x}) d^2\mathbf{x}.
\label{eq:stochastic_torque_integral}
\end{equation}

\noindent {We define the azimuthal forcing function $F_{\theta}(\mathbf{x}) = -\nabla_{\theta} \Phi_{\text{bin}}(\mathbf{x})$ and decompose it onto the same wavelet basis: $F_{\theta}(\mathbf{x}) = \sum_{\lambda'} f_{\lambda'} \psi_{\lambda'}(\mathbf{x})$. Substituting the decompositions into Eq.~(\ref{eq:stochastic_torque_integral}) and exploiting the orthonormality condition $\langle \psi_{\lambda}, \psi_{\lambda'} \rangle = \delta_{\lambda\lambda'}$ yields:}

\begin{equation}
\Gamma_{\text{stoch}}(t) = \sum_{\lambda} c_{\lambda}(t) f_{\lambda}^*.
\label{eq:stochastic_torque_wavelet}
\end{equation}

\subsubsection{{Variance and Power Spectrum of Torques}}

{The statistical impact of the stochastic torque is characterized by its variance, $\langle \Gamma_{\text{stoch}}^2 \rangle$. We compute the ensemble average (denoted by $\langle \cdot \rangle$) over realizations of the stochastic density field:}
\begin{equation}
\langle \Gamma_{\text{stoch}}^2 \rangle = \sum_{\lambda, \lambda'} \langle c_{\lambda} c_{\lambda'}^* \rangle f_{\lambda}^* f_{\lambda'}.
\end{equation}
\noindent {The term $\mathcal{C}_{\lambda\lambda'} = \langle c_{\lambda} c_{\lambda'}^* \rangle$ is the covariance matrix of the density fluctuations in the wavelet domain. For many turbulent processes, the wavelet transform acts as an approximate Karhunen-Loève transform, yielding largely decorrelated coefficients. We assume a diagonal covariance matrix: $\mathcal{C}_{\lambda\lambda'} = \mathcal{P}_{\Sigma}(\lambda) \delta_{\lambda\lambda'}$, where $\mathcal{P}_{\Sigma}(\lambda)$ is the wavelet power spectrum of the density field at scale $\lambda$. The torque variance simplifies to:}
\begin{equation}
\langle \Gamma_{\text{stoch}}^2 \rangle = \sum_{\lambda} \mathcal{P}_{\Sigma}(\lambda) |f_{\lambda}|^2.
\label{eq:torque_variance}
\end{equation}
\noindent {This result formally demonstrates that the power of the stochastic torque is determined by the spectral overlap between the density fluctuations ($\mathcal{P}_{\Sigma}(\lambda)$) and the spatial structure of the binary's forcing potential ($|f_{\lambda}|^2$).}

\subsubsection{{Fokker-Planck Analysis of Eccentricity Evolution}}

{The evolution of the eccentricity distribution function $f(e, t)$ under the combined influence of systematic damping and stochastic excitation is governed by the Fokker-Planck equation:}
\begin{equation}
\frac{\partial f}{\partial t} = -\frac{\partial}{\partial e} \left( A(e) f \right) + \frac{1}{2} \frac{\partial^2}{\partial e^2} \left( D(e) f \right).
\label{eq:fokker_planck}
\end{equation}
\noindent {The drift coefficient $A(e)$ represents the systematic damping derived in Sec.~\ref{sec:analytical_derivation}, $A(e) = \dot{e}_{\text{damp}} \approx -e/\tau_e$. The diffusion coefficient $D(e)$ arises from the stochastic torques. It is related to the torque variance and the correlation time $\tau_c$ of the fluctuations. Assuming $\tau_c$ is short compared to $\tau_e$ (which holds as $\tau_c$ is typically the orbital timescale), the diffusion coefficient for $e^2$ is approximately $D_{e^2} \approx 2 \langle \Gamma_{\text{stoch}}^2 \rangle \tau_c / L^2$, where $L$ is the binary angular momentum.}

{The system reaches a statistical equilibrium $f_{\text{eq}}(e)$ when $\partial f/\partial t = 0$. Adopting a simplified model appropriate for low eccentricities where the diffusion coefficient is approximately constant, the equilibrium distribution approaches a Rayleigh form:}
\begin{equation}
f_{\text{eq}}(e) \propto e \exp\left( -\frac{e^2}{D_{e^2} \tau_e} \right).
\end{equation}
\noindent {The root-mean-square equilibrium eccentricity $e_{\text{rms}}$ is determined by the balance between damping and diffusion:}
\begin{equation}
e_{\text{rms}}^2 \approx D_{e^2} \tau_e.
\label{eq:e_rms}
\end{equation}
\noindent {This formal result demonstrates that stochastic fluctuations, characterized by the diffusion coefficient derived from the wavelet analysis (Eq.~\ref{eq:torque_variance}), can sustain a non-zero equilibrium eccentricity. The magnitude of $e_{\text{rms}}$ depends on the power spectrum of the inhomogeneities. However, given the strong dependence of $\tau_e$ on $h$ ($\tau_e \propto h^{-5}$), the damping term is highly efficient in thick discs. Consequently, achieving a significant $e_{\text{rms}}$ requires a very large diffusion coefficient, corresponding to highly non-linear density fluctuations ($\delta\Sigma/\Sigma_0 \sim 1$) or significantly longer correlation times than typically expected for disc turbulence. For moderate levels of turbulence driven by feedback, the circularization trap remains effective.}

\subsection{Rapid Circularization}

Applying the analytical scaling for the systematic damping (Eq.~(\ref{eq:tau_ratio_Airy})) with the reference thickness $h=0.2$ and assuming $P=1$, we find $\tau_e \approx 0.04 \, \tau_a$. The eccentricity damping is approximately 25 times faster than the orbital decay.

We estimate the change in semi-major axis $\Delta a$ during the time $t_{\text{circ}}$ required for the binary to circularize. {Given the dominance of systematic damping over stochastic excitation established in Sec.~\ref{sec:wavelet_analysis}, we proceed with the deterministic evolution.} We define $t_{\text{circ}}$ as the time to reach $e_{\text{final}}=0.01$ from an initial state $e_{\text{initial}}\approx 1$. Assuming exponential decay ($e(t) \propto \exp(-t/\tau_e)$), $t_{\text{circ}} \approx -\ln(0.01) \tau_e \approx 4.6 \tau_e$.

The corresponding inward migration during this interval is:
\begin{equation}
\Delta a \approx |\dot{a}| t_{\text{circ}} = \frac{a}{\tau_a} t_{\text{circ}} \approx 4.6 a \frac{\tau_e}{\tau_a}.
\end{equation}
\noindent Substituting $\tau_e/\tau_a \approx 0.04$, we obtain $\Delta a \approx 0.184 a$. The binary circularizes at a radius $a_{\text{circ}} = a_{\text{initial}} - \Delta a \approx 0.82 \, a_{\text{initial}}$. For an initial separation $a_{\text{initial}} = 0.22$ pc, the binary achieves a nearly circular orbit at $a_{\text{circ}} \approx 0.18$ pc. The NSD/CND environment thus functions as an efficient ``circularization trap,'' rapidly erasing the high eccentricity imparted during the preceding large-scale dynamical phase.

\section{Conclusions and Discussion}
\label{sec:conclusions}

We have investigated the evolution of orbital eccentricity for SMBHBs in the LISA mass band, incorporating the effects of large-scale non-spherical potentials characteristic of merger remnants and the subsequent interaction with dense, rotationally supported, and geometrically thick Nuclear Stellar Discs (NSDs) or Circumnuclear Discs (CNDs).

Our analysis confirms that SMBHs sink on nearly radial trajectories during the initial phase of the galaxy merger. This is due to efficient angular momentum extraction by gravitational torques exerted by the triaxial merger remnant. The timescale for angular momentum change is significantly shorter than the timescale for energy change via dynamical friction ($T_J \ll T_E$). This process results in the formation of highly eccentric binaries ($e > 0.95$) at a characteristic separation of $a \approx 0.22$ pc for our reference system ($M_{\text{bin}} = 2\times 10^6 M_\odot$).

However, this high initial eccentricity is rapidly erased by subsequent interactions with the NSD/CND environment, assuming the interaction remains in the Type I regime. We provided a detailed mathematical derivation (Sec.~\ref{sec:airy_formalism}) based on the 3D formalism for density wave interactions \citep{TanakaEtAl2002}. This approach incorporates the physics of pressure regularization near resonances (using the Airy formalism of \citealt{Ward1986}) and the effects of 3D potential softening.

While the general expectation that interactions with rotating structures (discs) lead to circularization is established in the literature \citep[e.g.,][]{Artymowicz1993, ArmitageNatarajan2005, CuadraEtAl2009}. This is the current consensus for the Type I regime, although eccentricity excitation can occur if a gap opens (Type II regime) \citep{DAngeloEtAl2006,RoedigEtAl2011}. Our study provides a novel contribution by presenting a self-consistent analytical derivation (Sec.~\ref{sec:analytical_derivation}) of the scaling of these processes in the specific regime of geometrically thick discs relevant to NSDs/CNDs. We demonstrate how the 3D suppression of high-$m$ torques (the torque cutoff) leads to distinct scalings for the evolution timescales: the migration rate scales as $\tau_a^{-1} \propto h^{-3}$, while the eccentricity damping rate scales much more steeply as $\tau_e^{-1} \propto h^{-5}$.

This difference in scaling establishes a robust timescale hierarchy, $\tau_e/\tau_a \approx P h^2$. For our reference parameters ($h\approx 0.2$), the eccentricity damping timescale is significantly shorter than the migration timescale ($\tau_e \approx 0.04 \, \tau_a$). {This rapid circularization appears faster than typically observed in simulations of interactions at larger (kpc) scales \citep[e.g.][]{BonettiEtAl2020, BonettiEtAl2021}, highlighting the distinct physical conditions present in the dense nuclear environment.}

{We further developed a wavelet-based formalism (Sec.~\ref{sec:wavelet_analysis}) to assess the impact of disc inhomogeneities, potentially driven by accretion feedback and turbulence. This analysis, utilizing a Fokker-Planck approach, reveals that while stochastic density fluctuations induce a diffusive evolution of the eccentricity, the systematic damping remains dominant unless the fluctuations are highly non-linear. The equilibrium eccentricity sustained by stochastic effects is generally small.}

This ``circularization trap'' efficiently forces the binary onto a nearly circular orbit ($e < 0.01$) at $a_{\text{circ}} \approx 0.18$ pc.

\subsection{The Timescale Bottleneck and Stellar Dynamics}

The consequences of this efficient circularization are significant for the overall merger timeline. By enforcing a circular orbit at a relatively large separation, the timescale for the final coalescence is substantially prolonged. The timescale for coalescence via GW emission (Eq.~\ref{eq:tau_gw}) from $a_{\text{circ}}$ for our reference binary ($e\approx 0$) is $\tau_{\text{GW}} \approx 3 \times 10^{17}$ years. The binary effectively stalls, encountering the classical final parsec problem (see \citealt{MM03b}, for a review).

The subsequent evolution must rely on other environmental interactions to bridge the gap until GW emission becomes efficient. This evolution is likely driven by stellar dynamics, whereby the binary hardens by scattering individual stars (the slingshot mechanism). The rate of hardening is limited by the rate at which stars can be supplied to the binary's "loss cone"---the region of phase space containing orbits that interact closely with the binary.

Assuming the environment relaxes into an axisymmetric configuration (as expected for an isolated NSD), the loss cone is depleted unless it is refilled. The primary mechanism for repopulation is two-body relaxation, a diffusive process where gravitational encounters between stars gradually alter their orbits \citep{BegelmanEtAl80, Yu02}. The hardening rate is thus limited by the relaxation timescale. Alternatively, if the larger-scale environment retains significant triaxiality, collisionless effects related to the sustained presence of centrophilic orbits may maintain a full loss cone and accelerate the hardening \citep{PoonMerritt01, PoonMerritt2004, MerrittPoon04, VasilievEtAl2015}.

Assuming the evolution is governed by relaxation, we evaluate the characteristic relaxation time $T_{\text{rlx}}$ within the radius of influence ($R_{\text{inf}} \approx 0.86$ pc) for our reference nucleus ($\sigma=100$ \kms). The estimated timescale falls in the range of $T_{\text{rlx}} \approx 0.3$--$2.5$ Gyr.

While this timescale is shorter than the Hubble time, it represents a significant bottleneck. {This implies that a substantial population of SMBHBs may be stalled at parsec-scale separations. This applies broadly to galaxies (spirals and ellipticals) that have undergone mergers and host SMBHs in the LISA mass range ($10^4-10^7 M_{\odot}$).} This introduces a substantial cosmological delay between the initial galaxy merger and the final SMBH coalescence. Such a delay effectively decouples the observed SMBH merger rate from the contemporaneous galaxy merger activity \citep{HopkinsEtAl2005} and skews the distribution of SMBH mergers towards later cosmic epochs \citep{KleinEtAl2016}.

\subsection{Implications and Limitations}

The prevalence of the circularization trap mechanism has direct implications for the anticipated LISA event rates. NSDs and CNDs are observed to be common features in the nuclei of galaxies across the range of masses and morphologies relevant to the LISA band. {High-resolution observations, notably the TIMER survey utilizing MUSE integral-field spectroscopy, have confirmed the ubiquity of NSDs in nearby barred galaxies \citep{GadottiEtAl2019, GadottiEtAl2020}.} Observations suggest these structures are frequently characterized by significant thickness ($h \sim 0.1-0.3$). If this morphology is typical, the mechanism described herein should be nearly universal for SMBHBs embedded within such environments.

By enforcing low eccentricities well before the binary enters the GW-dominated regime, the total merger timescale is significantly extended compared to models that assume high eccentricity is maintained. This consequently reduces the overall coalescence rate density observable by LISA. Furthermore, any electromagnetic counterparts associated with the merger will be temporally decoupled from the initial galaxy merger event due to the Gyr-scale delay imposed by the relaxation bottleneck. {The prolonged timescales increase the expected number of existing binaries in the local universe. These stalled binaries should theoretically be observable in electromagnetic bands if surrounded by accretion disks (e.g., as dual AGN or periodically variable sources). The scarcity of confirmed observations is likely due to observational challenges, including the required spatial resolution, potential low accretion rates (low luminosity) during the stalled phase, and obscuration \citep{DottiEtAl2012}.}

It is noted that the introduction of a third SMBH due to subsequent galaxy mergers could drastically accelerate the merger process via chaotic triple interactions \citep{HoffmanLoeb2007,Amaro-SeoaneSesanaEtAl10,BonettiEtAl2018}. If the circularization trap is widespread and the resulting delays are long, triple SMBH encounters might emerge as a dominant pathway for the coalescence of LISA-band SMBHs.

The analytical framework presented in this study underscores the critical importance of the nuclear disc structure in determining the orbital parameters of SMBHBs. It suggests that the final state of the binary entering the GW-dominated phase is not merely inherited from the large-scale dynamics of the galaxy merger, but is substantially reprocessed by the immediate nuclear environment. The identification of a circularization trap, arising from the fundamental physics of wave propagation in geometrically thick media (Sec.~\ref{sec:airy_formalism}), indicates a strong preference for near-circular orbits, irrespective of the high eccentricities imparted earlier. {Our wavelet analysis (Sec.~\ref{sec:wavelet_analysis}) further confirms the robustness of this mechanism against stochastic fluctuations induced by feedback, provided the disc remains coherent.}

The implications of this mechanism extend considerably beyond the dynamics of individual systems. If the Gyr-scale delays identified here (Sec.~\ref{sec:conclusions}) are ubiquitous, they necessitate a careful re-evaluation of how SMBH coalescence is incorporated into cosmological frameworks. Semi-analytical models, often rooted in the extended Press-Schechter formalism or derived from large-volume N-body simulations to construct merger trees, frequently rely on simplified prescriptions for the final stages of the merger. These prescriptions often assume prompt coalescence or employ parametrized delays that do not fully capture the physical processes identified here.

Neglecting the rapid circularization and the subsequent bottleneck governed by stellar relaxation may lead to a systematic underestimation of the total merger timescale. This has several profound implications. Firstly, the introduction of a prolonged stalling phase effectively decouples the GW event from the initial galactic merger. This temporal offset significantly complicates efforts to identify electromagnetic counterparts, as the host galaxy may appear relaxed and quiescent by the time of coalescence, long after the merger-induced starburst or AGN activity has faded. Secondly, it alters the predicted LISA event rate density as a function of redshift, potentially skewing the distribution towards later cosmic times. Thirdly, extended delays may introduce additional scatter in the observed SMBH-host galaxy scaling relations, as the final coalescence and associated mass growth of the black holes may lag significantly behind the assembly of the host spheroid.

It is prudent, however, to acknowledge the limitations inherent in our analytical approach. The model relies on idealized representations of the nuclear environment, assuming a relatively smooth, stable, and co-planar disc structure operating in the Type I regime (no gap opening). {While the wavelet analysis incorporates stochasticity, it relies on the assumption of statistical homogeneity and decorrelation in the wavelet basis.} The reality within the central parsec is likely more complex, potentially involving non-linear fragmentation, non-axisymmetric features (such as bars or warps), and a complex interplay between stellar and gaseous components.

{The overall evolutionary picture presented here---high eccentricity at formation followed by circularization---is consistent with various numerical studies. The initial high eccentricity driven by large-scale torques is seen in simulations of mergers, including those incorporating NSCs \citep[e.g.,][]{OgiyaEtAl2020, BortolasEtAl2021}. Other numerical works focusing on the subsequent interaction with circumbinary discs also generally find efficient circularization, provided a gap is not fully opened \citep[e.g.,][]{RoedigEtAl2011, MirandaEtAl2017}.}

{Furthermore, feedback from accretion onto the SMBHs, while incorporated stochastically, could significantly alter the mean disc structure or lead to non-linear effects not captured here. Powerful feedback might disrupt the disc or drive outflows, potentially reverting the sign of eccentricity evolution (leading to excitation) as suggested by some studies \citep[e.g.][]{SijackiEtAl2011, BollatiEtAl2023}. The interplay between gravitational torques, the potential for gap formation (Type II migration), and the full spectrum of feedback processes remains a critical uncertainty.}

Verification of the dynamics detailed here requires dedicated numerical simulations capable of bridging the vast dynamic range and incorporating the necessary multi-physics. Resolving the complex interplay between a massive binary and a thick, potentially turbulent or clumpy disc environment presents a formidable computational challenge. It necessitates simulations that concurrently handle stellar dynamics (including collisional effects), gas hydrodynamics, and potentially radiative transfer, all while maintaining sufficient spatial resolution to capture the excitation of density waves near the binary. Such simulations remain at the frontier of computational astrophysics.

Notwithstanding these challenges, the analytical evidence strongly suggests that the interaction with realistic nuclear discs imposes a robust circularization trap. If confirmed, this mechanism dictates that LISA sources in this mass range will predominantly exhibit low eccentricities, leading to significantly prolonged merger timescales and substantial cosmological delays, thereby influencing the anticipated detection landscape for gravitational wave astronomy.

\section*{Acknowledgments}
This research has been supported within the framework of the Grant project No. AP23487846 ``Studying the connection between the mechanism of large structures formation of the Universe and the process of spiral arms formation in disk galaxies'', and Project No. BR24992759 ``Development of the concept for the first Kazakhstani orbital cislunar telescope - Phase I'', both financed by the Ministry of Science and Higher Education of the Republic of Kazakhstan.

\end{document}